\documentclass{sigchi}

\toappear{%
Permission to make digital or hard copies of all or part of this work for personal or classroom use is granted without fee provided that copies are not made or distributed for profit or commercial advantage and that copies bear this notice and the full citation on the first page. Copyrights for components of this work owned by others than the author(s) must be honored. Abstracting with credit is permitted. To copy otherwise, or republish, to post on servers or to redistribute to lists, requires prior specific permission and/or a fee. Request permissions from \href{mailto:Permissions@acm.org}{Permissions@acm.org}. \\
\emph{CHI'16}, May 07--May 12, 2016, San Jose, CA, USA \\
Copyright is held by the owner/author(s). Publication rights licensed to ACM \\
ACM 978-1-4503-3362-7/16/05\ldots \$15.00 \\
DOI\@: \url{http://dx.doi.org/10.1145/2858036.2858115}
}

\usepackage{balance}  
\usepackage{graphics} 
\usepackage{txfonts}
\usepackage{times}    
\usepackage[pdflang={en-US},pdftex]{hyperref}
\usepackage[usenames,dvipsnames]{xcolor}
\usepackage{textcomp}
\usepackage{booktabs}
\usepackage{ccicons}
\usepackage{todonotes}
\usepackage{dcolumn}
\usepackage{multirow}
\usepackage{caption}
\usepackage{subfigure}

\def\plaintitle{Embracing Error to Enable Rapid Crowdsourcing}
\def\plainauthor{Ranjay Krishna, Kenji Hata, Stephanie Chen, Joshua Kravitz, David A. Shamma, Li Fei-Fei, Michael S. Bernstein}

\def\plainkeywords{Human computation; Crowdsourcing; RSVP}

\makeatletter
\def\url@leostyle{%
  \@ifundefined{selectfont}{\def\UrlFont{\sf}}{\def\UrlFont{\small\bf\ttfamily}}}
\makeatother
\def\pprw{8.5in}
\def\pprh{11in}

\definecolor{linkColor}{RGB}{6,125,233}
\hypersetup{%
  pdftitle={\plaintitle},
  pdfauthor={\plainauthor},
  pdfkeywords={\plainkeywords},
  pdfdisplaydoctitle=true, 
  bookmarksnumbered,
  pdfstartview={FitH},
  colorlinks,
  citecolor=black,
  filecolor=black,
  linkcolor=black,
  urlcolor=linkColor,
  breaklinks=true,
  hypertexnames=false
}

\begin{document}

\title{\plaintitle}

\numberofauthors{1}
\author{%
  \alignauthor{Ranjay Krishna\textsuperscript{1}, Kenji Hata\textsuperscript{1}, Stephanie Chen\textsuperscript{1}, Joshua Kravitz\textsuperscript{1},\\ David A. Shamma\textsuperscript{2}, Li Fei-Fei\textsuperscript{1}, Michael S. Bernstein\textsuperscript{1} \\
    \affaddr{Stanford University\textsuperscript{1}, Yahoo! Labs\textsuperscript{2}}\\
    \email{\{ranjaykrishna, kenjihata, stephchen, kravitzj, feifeili, msb\}@cs.stanford.edu, aymans@acm.org}}\\
}


\maketitle

\begin{abstract}
Microtask crowdsourcing has enabled dataset advances in social science and machine learning, but existing crowdsourcing schemes are too expensive to scale up with the expanding volume of data. To scale and widen the applicability of crowdsourcing, we present a technique that produces extremely rapid judgments for binary and categorical labels. Rather than punishing all errors, which causes workers to proceed slowly and deliberately, our technique speeds up workers' judgments to the point where errors are acceptable and even expected. We demonstrate that it is possible to rectify these errors by randomizing task order and modeling response latency. We evaluate our technique on a breadth of common labeling tasks such as image verification, word similarity, sentiment analysis and topic classification. Where prior work typically achieves a 0.25$\times$ to 1$\times$ speedup over fixed majority vote, our approach often achieves an order of magnitude (10$\times$) speedup.
\end{abstract}


\category{H.5.m.}{Information interfaces and presentation (e.g., HCI)}{Miscellaneous.}

\section{Introduction}
Social science~\cite{kittur2008crowdsourcing, mason2012conducting}, interactive systems~\cite{fast2014emergent, kumar2013webzeitgeist} and machine learning~\cite{deng2009imagenet, lin2014microsoft} are becoming more and more reliant on large-scale, human-annotated data. Increasingly large annotated datasets have unlocked a string of social scientific insights~\cite{gilbert2009predicting, burke2013using} and machine learning performance improvements~\cite{krizhevsky2012imagenet, girshick2014rich, vinyals2014show}. One of the main enablers of this growth has been microtask crowdsourcing~\cite{snow2008cheap}. Microtask crowdsourcing marketplaces such as Amazon Mechanical Turk offer a scale and cost that makes such annotation feasible. As a result, companies are now using crowd work to complete hundreds of thousands of tasks per day~\cite{marcuswaran}.

However, even microtask crowdsourcing can be insufficiently scalable, and it remains too expensive for use in the production of many industry-size datasets~\cite{josephy2013crowdscale}. Cost is bound to the amount of work completed per minute of effort, and existing techniques for speeding up labeling (reducing the amount of required effort) are not scaling as quickly as the volume of data we are now producing that must be labeled~\cite{thomee2016yfcc100m}. To expand the applicability of crowdsourcing, 
the number of items annotated per minute of effort needs to increase substantially.

\begin{figure*}
  \centering
  \includegraphics[width=2\columnwidth]{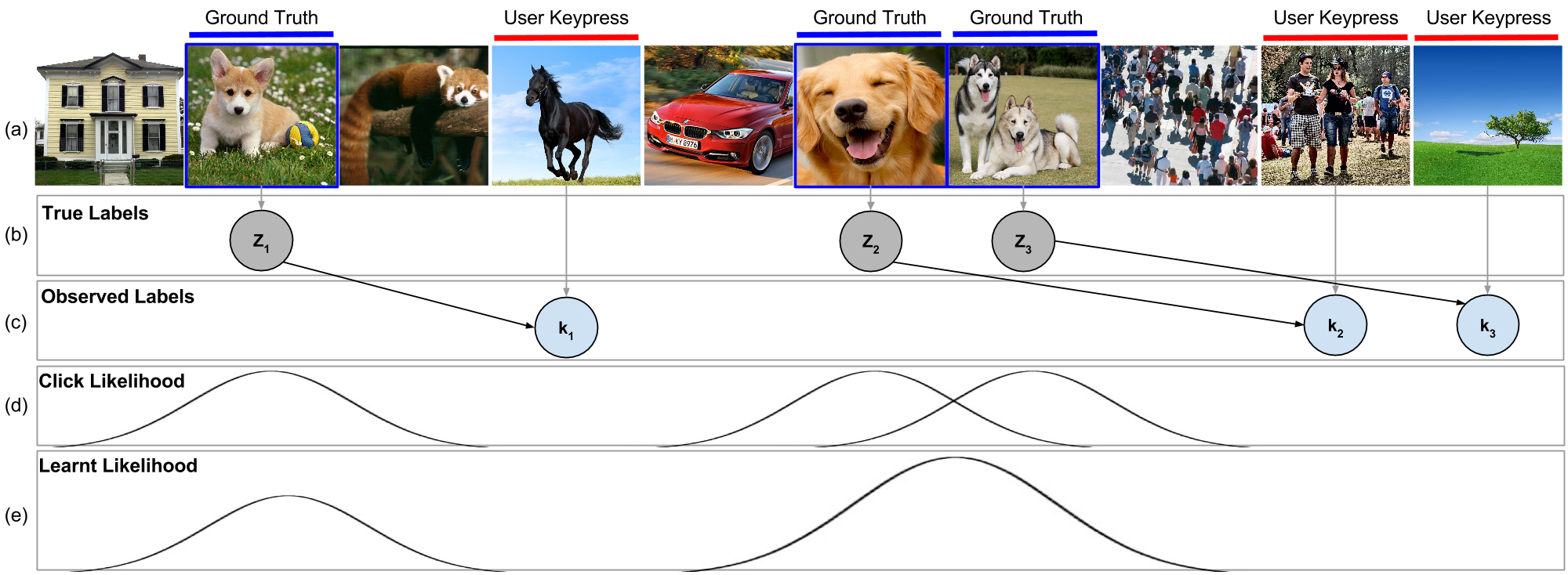}
  \caption{(a) Images are shown to workers at 100ms per image. Workers react whenever they see a dog. (b) The true labels are the ground truth dog images. (c) The workers' keypresses are slow and occur several images after the dog images have already passed. We record these keypresses as the observed labels. (d) Our technique models each keypress as a delayed Gaussian to predict (e) the probability of an image containing a dog from these observed labels.}
  \label{fig:pull_figure}
\end{figure*}

In this paper, we focus on one of the most common classes of crowdsourcing tasks \cite{ipeirotis2010analyzing}: binary annotation. These tasks are yes-or-no questions, typically identifying whether or not an input has a specific characteristic. Examples of these types of tasks are topic categorization (e.g., ``Is this article about finance?'')~\cite{schapire2000boostexter}, image classification (e.g., ``Is this a dog?'')~\cite{deng2009imagenet, lin2014microsoft, li2003detecting}, audio styles~\cite{seetharaman2014crowdsourcing} and emotion detection~\cite{li2003detecting} in songs (e.g., ``Is the music calm and soothing?''), word similarity (e.g., ``Are \textit{shipment} and \textit{cargo} synonyms?'')~\cite{miller1991contextual} and sentiment analysis (e.g.,\ ``Is this tweet positive?'')~\cite{pang2008opinion}. 


Previous methods have sped up binary classification tasks by minimizing worker error.
A central assumption behind this prior work has been that workers make errors because they are not trying hard enough (e.g.,\ ``a lack of expertise, dedication [or] interest''~\cite{sheng2008get}).
Platforms thus punish errors harshly, for example by denying payment. Current methods calculate the minimum redundancy necessary to be confident that errors have been removed~\cite{sheng2008get, smyth1994knowledge, smyth1995inferring}. These methods typically result in a 0.25$\times$ to 1$\times$ speedup beyond a fixed majority vote~\cite{peng2010decision,russakovsky2015best,sheng2008get,karger2014budget}.

We take the opposite position: that designing the task to encourage some error, or even make errors inevitable, can produce far greater speedups.
Because platforms strongly punish errors, workers carefully examine even straightforward tasks to make sure they do not represent edge cases~\cite{martin2014being, irani2013turkopticon}. The result is slow, deliberate work. 
We suggest that there are cases where we can encourage workers to move quickly by telling them that making some errors is acceptable.
Though individual worker accuracy decreases, we can recover from these mistakes post-hoc algorithmically (Figure~\ref{fig:pull_figure}).

We manifest this idea via a crowdsourcing technique in which workers label a rapidly advancing stream of inputs. Workers are given a binary question to answer, and they observe as the stream automatically advances via a method inspired by rapid serial visual presentation (RSVP)~\cite{li2002rapid, fei2007we}. Workers press a key whenever the answer is ``yes'' for one of the stream items. Because the stream is advancing rapidly, workers miss some items and have delayed responses. However, workers are reassured that the requester expects them to miss a few items. To recover the correct answers, the technique randomizes the item order for each worker and model workers' delays as a normal distribution whose variance depends on the stream's speed. For example, when labeling whether images have a ``barking dog'' in them, a self-paced worker on this task takes 1.7s per image on average. With our technique, workers are shown a stream at 100ms per image. The technique models the delays experienced at different input speeds and estimates the probability of intended labels from the key presses.

We evaluate our technique by comparing the total worker time necessary to achieve the same precision on an image labeling task as a standard setup with majority vote.
The standard approach takes three workers an average of 1.7s each for a total of 5.1s.
Our technique achieves identical precision (97\%) with five workers at 100ms each, for a total of 500ms of work. The result is an order of magnitude speedup of 10$\times$. 

This relative improvement is robust across both simple tasks, such as identifying dogs, and complicated tasks, such as identifying ``a person riding a motorcycle'' (interactions between two objects) or ``people eating breakfast'' (understanding relationships among many objects). We generalize our technique to other tasks such as word similarity detection, topic classification and sentiment analysis. Additionally, we extend our method to categorical classification tasks through a ranked cascade of binary classifications. Finally, we test workers' subjective mental workload and find no measurable increase.

\textbf{Contributions}. We make the following contributions:
\begin{enumerate}
  \setlength{\itemsep}{1pt}
\item We introduce a rapid crowdsourcing technique that makes errors normal and even inevitable. We show that it can be used to effectively label large datasets by achieving a speedup of an order of magnitude on several binary labeling crowdsourcing tasks.
\item We demonstrate that our technique can be generalized to multi-label categorical labeling tasks, combined independently with existing optimization techniques, and deployed without increasing worker mental workload.
\end{enumerate}

\section{Related Work}
The main motivation behind our work is to provide an environment where humans can make decisions quickly. We encourage a margin of human error in the interface that is then rectified by inferring the true labels algorithmically. In this section, we review prior work on crowdsourcing optimization and other methods for motivating contributions. Much of this work relies on artificial intelligence techniques: we complement this literature by changing the crowdsourcing interface rather than focusing on the underlying statistical model.

Our technique is inspired by rapid serial visual presentation (RSVP), a technique for consuming media rapidly by aligning it within the foveal region and advancing between items quickly~\cite{li2002rapid, fei2007we}. RSVP has already been proven to be effective at speeding up reading rates~\cite{wobbrock2002webthumb}. RSVP users can react to a single target image in a sequence of images even at 125ms per image with 75\% accuracy~\cite{potter1976short}. However, when trying to recognize concepts in images, RSVP only achieves an accuracy of 10\% at the same speed~\cite{potter1969recognition}. In our work, we integrate multiple workers' errors to successfully extract true labels.

Many previous papers have explored ways of modeling workers to remove bias or errors from ground truth labels~\cite{whitehill2009whose, welinder2010multidimensional, zhou2012learning, peng2010decision, ipeirotis2010quality}. For example, an unsupervised method for judging worker quality can be used as a prior to remove bias on binary verification labels~\cite{ipeirotis2010quality}. Individual workers can also be modeled as projections into an open space representing their skills in labeling a particular image~\cite{whitehill2009whose}. Workers may have unknown expertise that may in some cases prove adversarial to the task. Such adversarial workers can be detected by jointly learning the difficulty of labeling a particular datum along with the expertises of workers~\cite{welinder2010multidimensional}. Finally, a generative model can be used to model workers' skills by minimizing the entropy of the distribution over their labels and the unknown true labels~\cite{zhou2012learning}. We draw inspiration from this literature, calibrating our model using a similar generative approach to understand worker reaction times. We model each worker's reaction as a delayed Gaussian distribution.

In an effort to reduce cost, many previous papers have studied the tradeoffs between speed (cost) and accuracy on a wide range of tasks~\cite{wah2014similarity, branson2010visual, wah2011multiclass, russakovsky2014imagenet}. Some methods estimate human time with annotation accuracy to jointly model the errors in the annotation process~\cite{wah2014similarity, branson2010visual, wah2011multiclass}. Other methods vary both the labeling cost and annotation accuracy to calculate a tradeoff between the two~\cite{jain2013predicting, deng2014scalable}. Similarly, some crowdsourcing systems optimize a budget to measure confidence in worker annotations~\cite{karger2011budget, karger2014budget}. Models can also predict the redundancy of non-expert labels needed to match expert-level annotations~\cite{sheng2008get}. Just like these methods, we show that non-experts can use our technique and provide expert-quality annotations; we also compare our methods to the conventional majority-voting annotation scheme.

Another perspective on rapid crowdsourcing is to return results in real time, often by using a retainer model to recall workers quickly~\cite{bernstein2011crowds, lasecki2011real,laput2015zensors}. Like our approach, real-time crowdsourcing can use algorithmic solutions to combine multiple in-progress contributions~\cite{lasecki2012real}. These systems' techniques could be fused with ours to create crowds that can react to bursty requests. 

\begin{figure*}
  \centering
  \includegraphics[width=2\columnwidth]{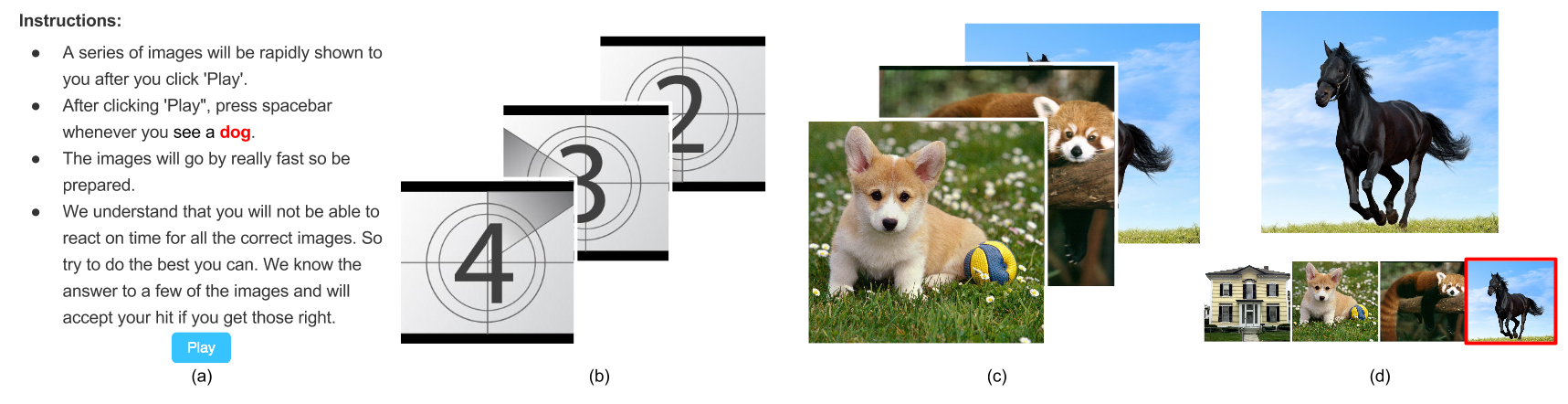}
  \caption{(a) Task instructions inform workers that we expect them to make mistakes since the items will be displayed rapidly. (b) A string of countdown images prepares them for the rate at which items will be displayed. (c) An example image of a ``dog'' shown in the stream---the two images appearing behind it are included for clarity but are not displayed to workers. (d) When the worker presses a key, we show the last four images below the stream of images to indicate which images might have just been labeled.}~\label{fig:interface}
\end{figure*}

One common method for optimizing crowdsourcing is active learning, which involves learning algorithms that interactively query the user. Examples include training image recognition~\cite{song2011contextualizing} and attribution recognition~\cite{parkash2012attributes} with fewer examples. 
Comparative models for ranking attribute models have also optimized crowdsourcing using active learning~\cite{liang2014beyond}. Similar techniques have explored optimization of the ``crowd kernel'' by adaptively choosing the next questions asked of the crowd in order to build a similarity matrix between a given set of data points~\cite{tamuz2011adaptively}.
Active learning needs to decide on a new task after each new piece of data is gathered from the crowd. Such models tend to be quite expensive to compute. Other methods have been proposed to decide on a set of tasks instead of just one task~\cite{vijayanarasimhan2010far}. We draw on this literature: in our technique, after all the images have been seen by at least one worker, we use active learning to decide the next set of tasks. We determine which images to discard and which images to group together and send this set to another worker to gather more information.


Finally, there is a group of techniques that attempt to optimize label collection by reducing the number of questions that must be answered by the crowd. For example, a hierarchy in label distribution can reduce the annotation search space~\cite{deng2014scalable}, and information gain can reduce the number of labels necessary to build large taxonomies using a crowd~\cite{chilton2013cascade, bragg2013crowdsourcing}. Methods have also been proposed to maximize accuracy of object localization in images~\cite{su2012crowdsourcing} and videos~\cite{vondrick2013efficiently}. Previous labels can also be used as a prior to optimize acquisition of new types of annotations~\cite{branson2014active}. One of the benefits of our technique is that it can be used independently of these others to jointly improve crowdsourcing schemes. We demonstrate the gains of such a combination in our evaluation.

\section{Error-Embracing Crowdsourcing}
\label{sec:interface_design}
Current microtask crowdsourcing platforms like Amazon Mechanical Turk incentivize workers to avoid rejections~\cite{irani2013turkopticon, martin2014being}, resulting in slow and meticulous work. But is such careful work necessary to build an accurate dataset? In this section, we detail our technique for rapid crowdsourcing by encouraging less accurate work.

The design space of such techniques must consider which tradeoffs are acceptable to make. The first relevant dimension is accuracy. When labeling a large dataset (e.g., building a dataset of ten thousand articles about housing), \textit{precision} is often the highest priority: articles labeled as on-topic by the system must in fact be about housing. \textit{Recall}, on the other hand, is often less important, because there is typically a large amount of available unlabeled data: even if the system misses some on-topic articles, the system can label more items until it reaches the desired dataset size. We thus develop an approach for producing high precision at high speed, sacrificing some recall if necessary.

The second design dimension involves the task characteristics. Many large-scale crowdsourcing tasks involve closed-ended responses such as binary or categorical classifications. These tasks have two useful properties. First, they are time-bound by users' perception and cognition speed rather than motor (e.g., pointing, typing) speed~\cite{cheng2015measuring}, since acting requires only a single button press. Second, it is possible to aggregate responses automatically, for example with majority vote. Open-ended crowdsourcing tasks such as writing~\cite{bernstein2010soylent} or transcription are often time-bound by data entry motor speeds and cannot be automatically aggregated. Thus, with our technique, we focus on closed-ended tasks.

\subsection{Rapid crowdsourcing of binary decision tasks}
Binary questions are one of the most common classes of crowdsourcing tasks. Each yes-or-no question gathers a label on whether each item has a certain characteristic. In our technique, rather than letting workers focus on each item too carefully, we display each item for a specific period of time before moving on to the next one in a rapid slideshow. For example, in the context of an image verification task, we show workers a stream of images and ask them to press the spacebar whenever they see a specific class of image. In the example in Figure~\ref{fig:interface}, we ask them to react whenever they see a ``dog.''

The main parameter in this approach is the length of time each item is visible. To determine the best option, we begin by allowing workers to work at their own pace. This establishes an initial average time period, which we then slowly decrease in successive versions until workers start making mistakes~\cite{cheng2015measuring}. Once we have identified this error point, we can algorithmically model workers' latency and errors to extract the true labels.

To avoid stressing out workers, it is important that the task instructions convey the nature of the rapid task and the fact that we expect them to make some errors.  Workers are first shown a set of instructions (Figure~\ref{fig:interface}(a)) for the task. They are warned that reacting to every single correct image on time is not feasible and thus not expected. We also warn them that we have placed a small number of items in the set that we know to be positive items. These help us calibrate each worker's speed and also provide us with a mechanism to reject workers who do not react to any of the items.

Once workers start the stream (Figure~\ref{fig:interface}(b)), it is important to prepare them for pace of the task. We thus show a film-style countdown for the first few seconds that decrements to zero at the same interval as the main task. Without these countdown images, workers use up the first few seconds getting used to the pace and speed. Figure~\ref{fig:interface}(c) shows an example ``dog'' image that is displayed in front of the user. The dimensions of all items (images) shown are held constant to avoid having to adjust to larger or smaller visual ranges.

When items are displayed for less than 400ms, workers tend to react to all positive items with a delay. If the interface only reacts with a simple confirmation when workers press the spacebar, many workers worry that they are too late because another item is already on the screen. Our solution is to also briefly display the last four items previously shown when the spacebar is pressed, so that workers see the one they intended and also gather an intuition for how far back the model looks. For example, in Figure~\ref{fig:interface}(d), we show a worker pressing the spacebar on an image of a horse. We anticipate that the worker was probably delayed, and we display the last four items to acknowledge that we have recorded the keypress. We ask all workers to first complete a qualification task in which they receive feedback on how quickly we expect them to react. They pass the qualification task only if they achieve a recall of 0.6 and precision of 0.9 on a stream of 200 items with 25 positives. We measure precision as the fraction of worker reactions that were within 500ms of a positive cue.

\begin{figure*}
  \centering
  \includegraphics[width=2\columnwidth]{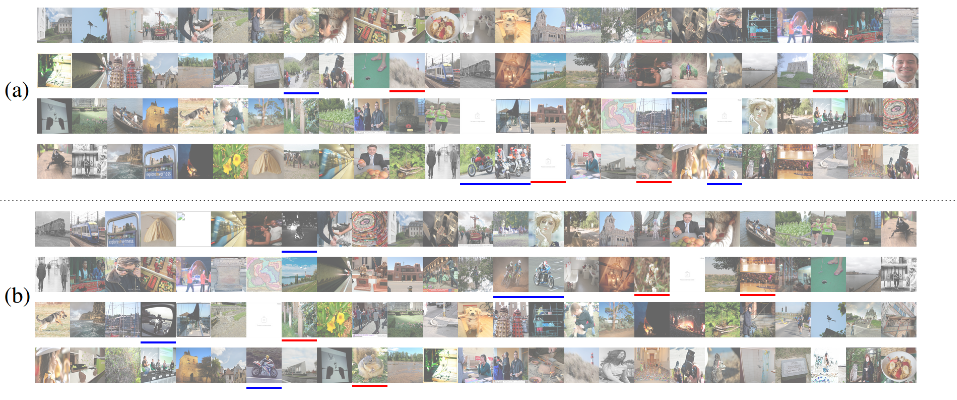}\label{fig:sub1}
    \caption{Example raw worker outputs from our interface. Each image was displayed for 100ms and workers were asked to react whenever they saw images of ``a person riding a motorcycle.'' Images are shown in the same order they appeared in for the worker. Positive images are shown with a blue bar below them and users' keypresses are shown as red bars below the image to which they reacted.}%
    \label{fig:qualitative_clicks}%
\end{figure*}

In Figure~\ref{fig:qualitative_clicks}, we show two sample outputs from our interface. Workers were shown images for 100ms each. They were asked to press the spacebar whenever they saw an image of ``a person riding a motorcycle.'' The images with blue bars underneath them are ground truth images of ``a person riding a motorcycle.'' The images with red bars show where workers reacted. The important element is that red labels are often delayed behind blue ground truth and occasionally missed entirely. Both Figures~\ref{fig:qualitative_clicks}(a) and~\ref{fig:qualitative_clicks}(b) have 100 images each with 5 correct images.

Because of workers' reaction delay, the data from one worker has considerable uncertainty. We thus show the same set of items to multiple workers in different random orders and collect independent sets of keypresses. This randomization will produce a cleaner signal in aggregate and later allow us to estimate the images to which each worker intended to react.

Given the speed of the images, workers are not able to detect every single positive image. For example, the last positive image in Figure~\ref{fig:qualitative_clicks}(a) and the first positive image in Figure~\ref{fig:qualitative_clicks}(b) are not detected. Previous work on RSVP found a phenomenon called ``attention blink''~\cite{broadbent1987detection}, in which a worker is momentarily blind to successive positive images. However, we find that even if two images of ``a person riding a motorcycle'' occur consecutively, workers are able to detect both and react twice (Figures~\ref{fig:qualitative_clicks}(a) and~\ref{fig:qualitative_clicks}(b)). 
If workers are forced to react in intervals of less than 400ms, though, the signal we extract is too noisy for our model to estimate the positive items.

\subsection{Multi-Class Classification for Categorical Data}
So far, we have described how rapid crowdsourcing can be used for binary verification tasks. Now we extend it to handle multi-class classification. Theoretically, all multi-class classification can be broken down into a series of binary verifications. For example, if there are $N$ classes, we can ask $N$ binary questions of whether an item is in each class. Given a list of items, we use our technique to classify them one class at a time. After every iteration, we remove all the positively classified items for a particular class. We use the rest of the items to detect the next class.

Assuming all the classes contain an equal number of items, the order in which we detect classes should not matter. A simple \textit{baseline approach} would choose a class at random and attempt to detect all items for that class first. However, if the distribution of items is not equal among classes, this method would be inefficient. Consider the case where we are trying to classify items into 10 classes, and one class has 1000 items while all other classes have 10 items. In the worst case, if we classify the class with 1000 examples last, those 1000 images would go through our interface 10 times (once for every class). Instead, if we had detected the large class first, we would be able to classify those 1000 images and they would only go through our interface once. With this intuition, we propose a \textit{class-optimized approach} that classifies the most common class of items first. We maximize the number of items we classify at every iteration, reducing the total number of binary verifications required.

\section{Model}
To translate workers' delayed and potentially erroneous actions into identifications of the positive items, we need to model their behavior. We do this by calculating the probability that a particular item is in the positive class given that the user reacted a given period after the item was displayed. By combining these probabilities across several workers with different random orders of the same images, these probabilities sum up to identify the correct items.

We use maximum likelihood estimation to predict the probability of an item being a positive example. Given a set of items $\mathcal{I} = \{I_1, \ldots, I_n\}$, we send them to $W$ workers in a different random order for each. From each worker $w$, we collect a set of keypresses $\mathcal{C}^{w} = \{c_1^w, \ldots, c_k^w\}$ where $w \in W$ and $k$ is the total number of keypresses from $w$. Our aim is to calculate the probability of a given item $P(I_i)$ being a positive example. Given that we collect keypresses from $W$ workers:
\setlength{\belowdisplayskip}{0pt} \setlength{\belowdisplayshortskip}{0pt}
\setlength{\abovedisplayskip}{0pt} \setlength{\abovedisplayshortskip}{0pt}
\vspace{5pt}
\begin{equation}
P(I_i) = \sum_{w} P(I_i | \mathcal{C}^{w}) P(\mathcal{C}^{w})
\end{equation}

where $P(\mathcal{C}) = \prod_{k} P(\mathcal{C}_k)$ is the probability of a particular set of items being keypresses. We set $P(C_k)$ to be constant, asssuming that it is equally likely that a worker might react to any item. Using Bayes' rule:
\vspace{0pt}
\begin{equation}
P(I_i | \mathcal{C}^{w}) = \frac{P(\mathcal{C}^{w} | I_i) P(I_i)}{P(\mathcal{C}^{w})}.
\end{equation}

$P(I_i)$ models our estimate of item $I_i$ being positive. It can be a constant, or it can be an estimate from a domain-specific machine learning algorithm~\cite{kamar2012combining}. For example, to calculate $P(I_i)$, if we were trying to scale up a dataset of ``dog'' images, we would use a small set of known ``dog'' images to train a binary classifier and use that to calculate $P(I_i)$ for all the unknown images. With image tasks, we use a pretrained convolutional neural network to extract image features~\cite{Simonyan14c} and train a linear support vector machine to calculate $P(I_i)$.

We model $P(\mathcal{C}^{w} | I_i)$ as a set of independent keypresses:

\begin{equation}
P(\mathcal{C}^{w} | I_i) = P(c_1^w, \ldots, c_k^w | I_i) = \prod_{k} P(\mathcal{C}^{w}_k | I_i).
\end{equation}

Finally, we model each keypress as a Gaussian distribution $\mathcal{N}(\mu, \sigma)$ given a positive item. We train the mean $\mu$ and variance $\sigma$ by running rapid crowdsourcing on a small set of items for which we already know the positive items. Here, the mean and variance of the distribution are modeled to estimate the delays that a worker makes when reacting to a positive item. 

Intuitively, the model works by treating each keypress as creating a Gaussian ``footprint'' of positive probability on the images about 400ms before the keypress (Figure~\ref{fig:pull_figure}). The model combines these probabilities across several workers to identify the images with the highest overall probability.

Now that we have a set of probabilities for each item, we need to decide which ones should be classified as positive. We order the set of items $\mathcal{I}$ according to likelihood of being in the positive class $P(I_i)$. We then set all items above a certain threshold as positive. This threshold is a hyperparameter that can be tuned to trade off precision vs.\ recall. 

In total, this model has two hyperparameters: (1) the threshold above which we classify images as positive and (2) the speed at which items are displayed to the user. We model both hyperparameters in a per-task (image verification, sentiment analysis, etc.) basis. For a new task, we first estimate how long it takes to label each item in the conventional setting with a small set of items. Next, we continuously reduce the time each item is displayed until we reach a point where the model is unable to achieve the same precision as the untimed case.


\begin{figure}[t]
\centering
  \includegraphics[width=0.9\columnwidth]{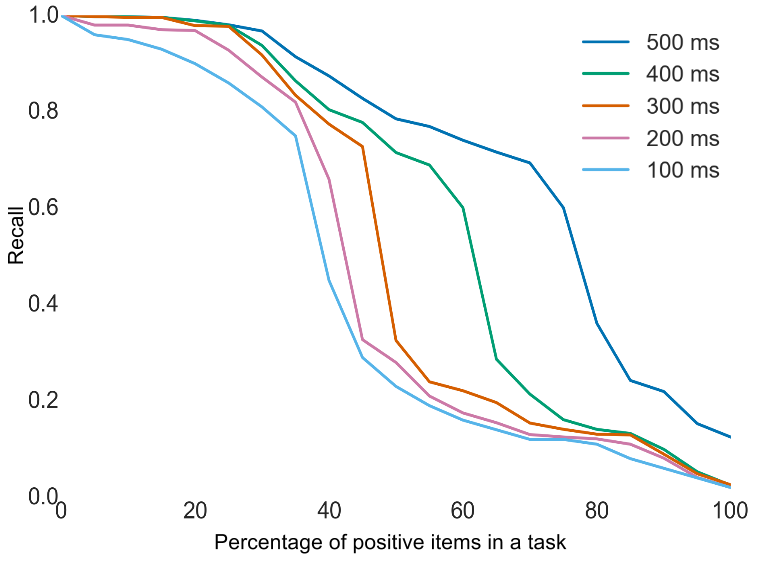}
  \caption{We plot the change in recall as we vary percentage of positive items in a task. We experiment at varying display speeds ranging from 100ms to 500ms. We find that recall is inversely proportional to the rate of positive stimuli and not to the percentage of positive items.}~\label{fig:recall_percentage}
\end{figure}

\begin{table*}[t]
\centering
\begin{tabular}{l l r r r r r r r}
  \textbf{Task} & & \multicolumn{3}{c}{\textbf{Conventional Approach}} & \multicolumn{3}{c}{\textbf{Our Technique}} & \textbf{Speedup} \\
  \cmidrule(r){3-5}
  \cmidrule(r){6-8}
  & 
  &
  \multicolumn{1}{c}{\textit{Time (s)}} & 
  \multicolumn{1}{c}{\textit{Precision}} & 
  \multicolumn{1}{c}{\textit{Recall}} & 
  \multicolumn{1}{c}{\textit{Time (s)}} & 
  \multicolumn{1}{c}{\textit{Precision}} & 
  \multicolumn{1}{c}{\textit{Recall}} & 
  \multicolumn{1}{c}{}\\
  \midrule
 \multirow{4}{*}{Image Verification} & Easy & 1.50 & 0.99 & 0.99 & 0.10 & 0.99 & 0.94 &  \textbf{9.00}$\times$ \\
                                   & Medium & 1.70 & 0.97 & 0.99 & 0.10 & 0.98 & 0.83 & \textbf{10.20}$\times$ \\
                                     & Hard & 1.90 & 0.93 & 0.89 & 0.10 & 0.90 & 0.74 & \textbf{11.40}$\times$ \\
                                     \cmidrule(r){2-9}
                                      & All Concepts & 1.70 & 0.97 & 0.96 & 0.10 & 0.97 & 0.81 & \textbf{10.20}$\times$ \\
 \midrule
 Sentiment Analysis & &  4.25 & 0.93 & 0.97 & 0.25 & 0.94 & 0.84 & \textbf{10.20}$\times$ \\
 Word Similarity    & &  6.23 & 0.89 & 0.94 & 0.60 & 0.88 & 0.86 &  \textbf{6.23}$\times$ \\
 Topic Detection    & & 14.33 & 0.96 & 0.94 & 2.00 & 0.95 & 0.81 & \textbf{10.75}$\times$ \\
 \bottomrule
\end{tabular}
\caption{We compare the conventional approach for binary verification tasks (image verification, sentiment analysis, word similarity and topic detection) with our technique and compute precision and recall scores. Precision scores, recall scores and speedups are calculated using 3 workers in the conventional setting. Image verification, sentiment analysis and word similarity used 5 workers using our technique, while topic detection used only 2 workers. We also show the time taken (in seconds) for 1 worker to do each task.}~\label{tab:precision_recall_speedup}
\end{table*}

\begin{figure*}[t]
  \centering
  \includegraphics[width=2\columnwidth]{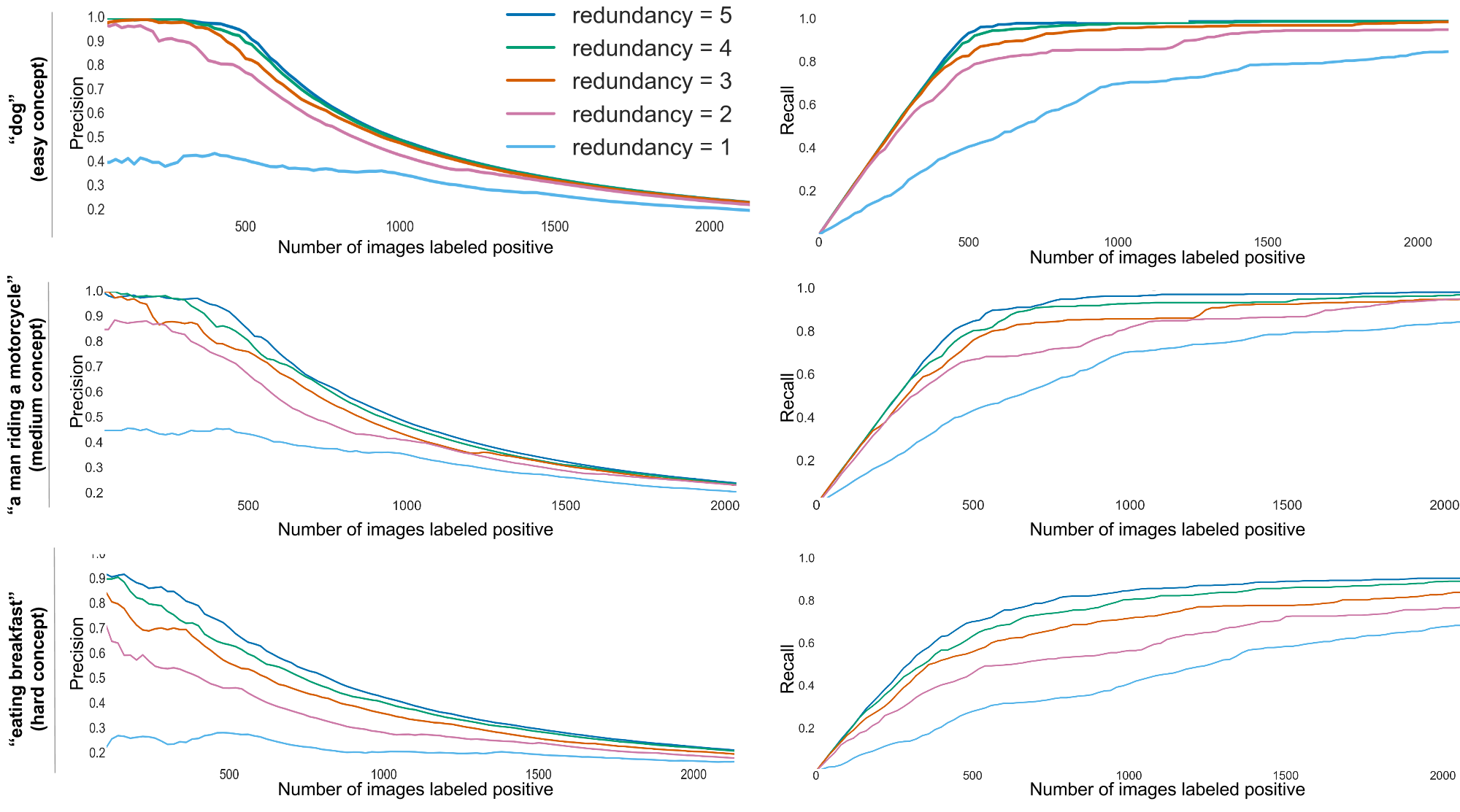}
  \caption{We study the precision (left) and recall (right) curves for detecting ``dog'' (top), ``a person on a motorcycle'' (middle) and ``eating breakfast'' (bottom) images with a redundancy ranging from 1 to 5. There are 500 ground truth positive images in each experiment. We find that our technique works for simple as well as hard concepts.}~\label{fig:preception_cognition}
\end{figure*}

\section{Calibration: Baseline Worker Reaction Time} 
Our technique hypothesizes that guiding workers to work quickly and make errors can lead to results that are faster yet with similar precision. We begin evaluating our technique by first studying worker reaction times as we vary the length of time for which each item is displayed. If worker reaction times have a low variance, we accurately model them. Existing work on RSVP estimated that humans usually react about 400ms after being presented with a cue~\cite{weichselgartner1987dynamics, reeves1986attention}. Similarly, the model human processor~\cite{card1983psychology} estimated that humans perceive, understand and react at least 240ms after a cue. We first measure worker reaction times, then analyze how frequently positive items can be displayed before workers are unable to react to them in time.

\textit{Method.} We recruited 1,000 workers on Amazon Mechanical Turk with 96\% approval rating and over 10,000 tasks submitted. Workers were asked to work on one task at a time. Each task contained a stream of 100 images of polka dot patterns of two different colors. Workers were asked to react by pressing the spacebar whenever they saw an image with polka dots of one of the two colors. Tasks could vary by two variables: the \textit{speed} at which images were displayed and the \textit{percentage} of the positively colored images. For a given task, we held the display speed constant. Across multiple tasks, we displayed images for 100ms to 500ms. We studied two variables: \textit{reaction time} and \textit{recall}. We measured the reaction time to the positive color across these speeds. To study recall (percentage of positively colored images detected by workers), we varied the ratio of positive images from 5\% to 95\%. We counted a keypress as a detection only if it occurred within 500ms of displaying a positively colored image.

\textit{Results.} Workers' reaction times corresponded well with estimates from previous studies. Workers tend to react an average of 378ms ($\sigma= 92$ms) after seeing a positive image. This consistency is an important result for our model because it assumes that workers have a consistent reaction delay.

As expected, recall is inversely proportional to the speed at which the images are shown. A worker is more likely to miss a positive image at very fast speeds. We also find that recall decreases as we increase the percentage of positive items in the task. To measure the effects of positive frequency on recall, we record the percentage threshold at which recall begins to drop significantly at different speeds and positive frequencies. From Figure~\ref{fig:recall_percentage}, at 100ms, we see that recall drops when the percentage of positive images is more than 35\%. As we increase the time for which an item is displayed, however, we notice that the drop in recall occurs at a much higher percentage. At 500ms, the recall drops at a threshold of 85\%. We thus infer that recall is inversely proportional to the \textit{rate} of positive stimuli and not to the percentage of positive images. From these results we conclude that at faster speeds, it is important to maintain a smaller percentage of positive images, while at slower speeds, the percentage of positive images has a lesser impact on recall. Quantitatively, to maintain a recall higher than 0.7, it is necessary to limit the frequency of positive cues to one every 400ms.

\section{Study 1: Image Verification}
In this study, we deploy our technique on image verification tasks and measure its speed relative to the conventional self-paced approach. Many crowdsourcing tasks in computer vision require verifying that a particular image contains a specific class or concept. We measure precision, recall and cost (in seconds) by the conventional approach and compare against our technique.

Some visual concepts are easier to detect than others. For example, detecting an image of a ``dog'' is a lot easier than detecting an image of ``a person riding a motorcycle'' or ``eating breakfast.'' While detecting a ``dog'' is a perceptual task, ``a person riding a motorcycle'' requires understanding of the interaction between the person and the motorcycle. Similarly, ``eating breakfast'' requires workers to fuse concepts of people eating a variety foods like eggs, cereal or pancakes. We test our technique on detecting three concepts: ``dog'' (easy concept), ``a person riding a motorcycle'' (medium concept) and ``eating breakfast'' (hard concept). In this study, we compare how workers fare on each of these three levels of concepts.

\textit{Method.} In this study, we compare the conventional approach with our technique on three (easy, medium and hard) concepts. We evaluate each of these comparisons using precision scores, recall scores and the speedup achieved. To test each of the three concepts, we labeled 10,000 images, where each concept had 500 examples. We divided the 10,000 images into streams of 100 images for each task. We paid workers \$0.17 to label a stream of 100 images (resulting in a wage of \$6 per hour~\cite{salehi2015we}). We hired over 1,000 workers for this study satisfying the same qualifications as the calibration task.

The conventional method of collecting binary labels is to present a crowd worker with a set of items. The worker proceeds to label each item, one at a time. Most datasets employ multiple workers to label each task because majority voting~\cite{snow2008cheap} has been shown to improve the quality of crowd annotations. These datasets usually use a redundancy of 3 to 5 workers~\cite{sheshadri2013square}. In all our experiments, we used a redundancy of 3 workers as our baseline. 

When launching tasks using our technique, we tuned the image display speed to 100ms. We used a redundancy of 5 workers when measuring precision and recall scores. To calculate \textit{speedup}, we compare the total worker time taken by all the 5 workers using our technique with the total worker time taken by the 3 workers using the conventional method. Additionally, we vary redundancy on all the concepts to from 1 to 10 workers to see its effects on precision and recall.


\textit{Results.} Self-paced workers take 1.70s on average to label each image with a concept in the conventional approach (Table~\ref{tab:precision_recall_speedup}). They are quicker at labeling the easy concept (1.50s per worker) while taking longer on the medium (1.70s) and hard (1.90s) concepts.

Using our technique, even with a redundancy of 5 workers, we achieve a speedup of 10.20$\times$ across all concepts. We achieve \textit{order of magnitude} speedups of $9.00\times$, $10.20\times$ and $11.40\times$ on the easy, medium and hard concepts. Overall, across all concepts, the precision and recall achieved by our technique is 0.97 and 0.81. Meanwhile the precision and recall of the conventional method is 0.97 and 0.96. We thus achieve the same precision as the conventional method. As expected, recall is lower because workers are not able to detect every single true positive example. As argued previously, lower recall can be an acceptable tradeoff when it is easy to find more unlabeled images.

Now, let's compare precision and recall scores between the three concepts. We show precision and recall scores in Figure~\ref{fig:preception_cognition} for the three concepts. Workers perform slightly better at finding ``dog'' images and find it the most difficult to detect the more challenging ``eating breakfast'' concept. With a redundancy of 5, the three concepts achieve a precision of 0.99, 0.98 and 0.90 respectively at a recall of 0.94, 0.83 and 0.74 (Table~\ref{tab:precision_recall_speedup}). The precision for these three concepts are identical to the conventional approach, while the recall scores are slightly lower. The recall for a more difficult cognitive concept (``eating breakfast'') is much lower, at 0.74, than for the other two concepts. More complex concepts usually tend to have a lot of contextual variance. For example, ``eating breakfast'' might include a person eating a ``banana,'' a ``bowl of cereal,'' ``waffles'' or ``eggs.'' We find that while some workers react to one variety of the concept (e.g.,\ ``bowl of cereal''), others react to another variety (e.g.,\ ``eggs''). 

\begin{figure*}
  \centering
  \includegraphics[width=2\columnwidth]{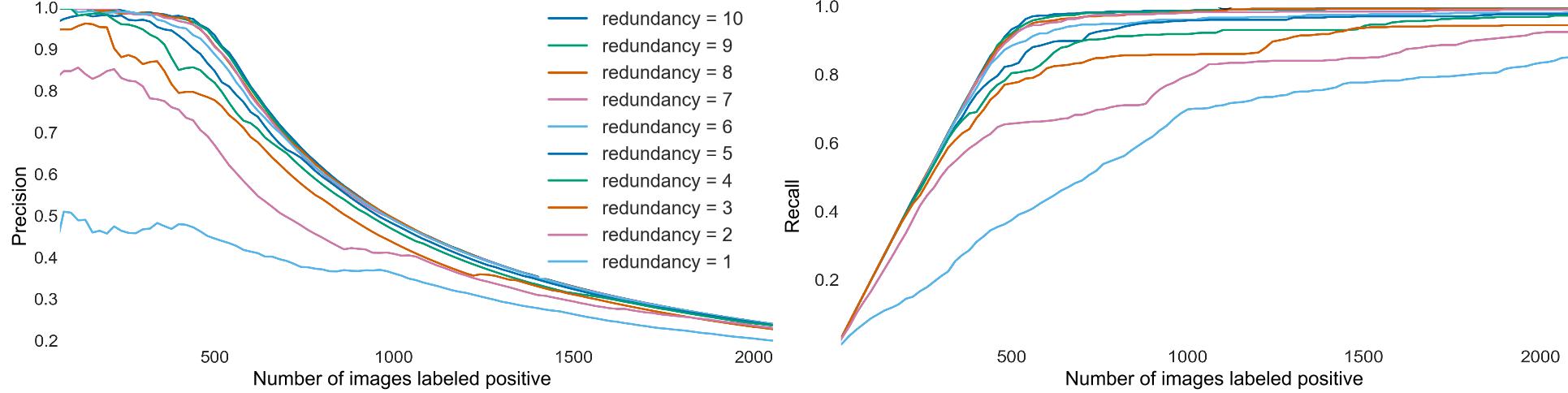}
  \caption{We study the effects of redundancy on recall by plotting precision and recall curves for detecting ``a person on a motorcycle'' images with a redundancy ranging from 1 to 10. We see diminishing increases in precision and recall as we increase redundancy. We manage to achieve the same precision and recall scores as the conventional approach with a redundancy of 10 while still achieving a speedup of $5\times$.}~\label{fig:redundancy_10}
\end{figure*}

When we increase the redundancy of workers to 10 (Figure~\ref{fig:redundancy_10}), our model is able to better approximate the positive images. We see diminishing increases in both recall and precision as redundancy increases. At a redundancy of 10, we increase recall to the same amount as the conventional approach (0.96), while maintaining a high precision (0.99) and still achieving a speedup of $5.1\times$.

We conclude from this study that our technique (with a redundancy of 5) can speed up image verification with easy, medium and hard concepts by an order of magnitude while still maintaining high precision. We also show that recall can be compensated by increasing redundancy.

\section{Study 2: Non-Visual Tasks}
So far, we have shown that rapid crowdsourcing can be used to collect image verification labels. We next test the technique on a variety of other common crowdsourcing tasks: sentiment analysis~\cite{pang2008opinion}, word similarity~\cite{snow2008cheap} and topic detection~\cite{LEWIS94d}.

\textit{Method.} In this study, we measure precision, recall and speedup achieved by our technique over the conventional approach. To determine the stream speed for each task, we followed the prescribed method of running trials and speeding up the stream until the model starts losing precision. For sentiment analysis, workers were shown a stream of tweets and asked to react whenever they saw a positive tweet. We displayed tweets at 250ms with a redundancy of 5 workers. For word similarity, workers were shown a word (e.g.,\ ``lad'') for which we wanted synonyms. They were then rapidly shown other words at 600ms and asked to react if they see a synonym (e.g.,\ ``boy''). Finally, for topic detection, we presented workers with a topic like ``housing'' or ``gas'' and presented articles of an average length of 105 words at a speed of 2s per article. They reacted whenever they saw an article containing the topic we were looking for. For all three of these tasks, we compare precision, recall and speed against the self-paced conventional approach with a redundancy of 3 workers. Every task, for both the conventional approach and our technique, contained 100 items. 

To measure the cognitive load on workers for labeling so many items at once, we ran the widely-used NASA Task Load Index (TLX)~\cite{colligan2015cognitive} on all tasks, including image verification. TLX measures the perceived workload of a task. We ran the survey on 100 workers who used the conventional approach and 100 workers who used our technique across all tasks.

\begin{figure*}[t]
  \centering
  \includegraphics[width=2\columnwidth]{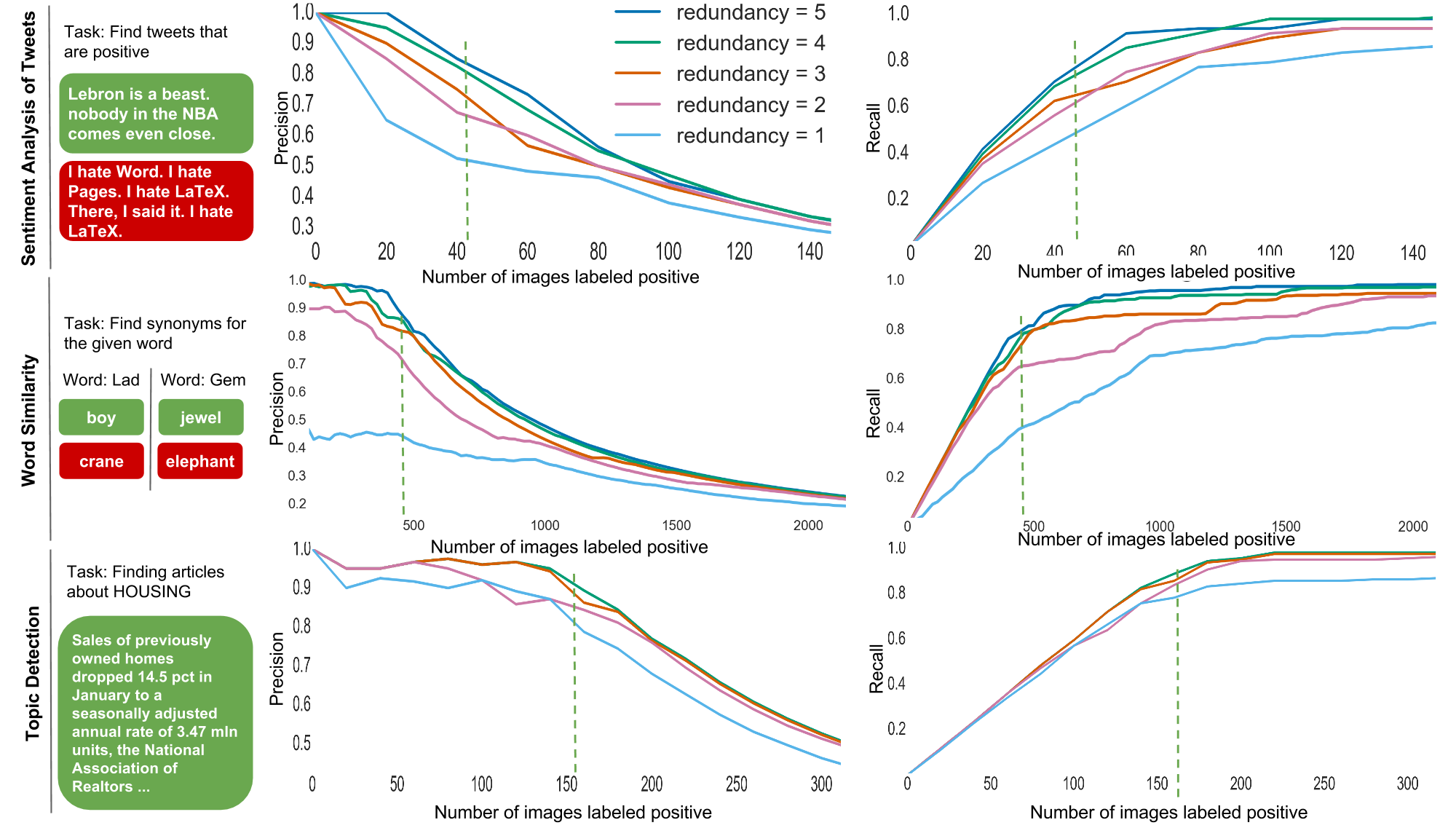}
  \caption{Precision (left) and recall (right) curves for sentiment analysis (top), word similarity (middle) and topic detection (bottom) images with a redundancy ranging from 1 to 5. Vertical lines indicate the number of ground truth positive examples.}~\label{fig:other_tasks}
\end{figure*}

\textit{Results.} We present our results in Table~\ref{tab:precision_recall_speedup} and Figure~\ref{fig:other_tasks}. For sentiment analysis, we find that workers in the conventional approach classify tweets in 4.25s. So, with a redundancy of 3 workers, the conventional approach would take 12.75s with a precision of 0.93. Using our method and a redundancy of 5 workers, we complete the task in 1250ms (250ms per worker per item) and 0.94 precision. Therefore, our technique achieves a speedup of $10.2\times$.

Likewise, for word similarity, workers take around 6.23s to complete the conventional task, while our technique succeeds at 600ms. We manage to capture a comparable precision of 0.88 using 5 workers against a precision of 0.89 in the conventional method with 3 workers. Since finding synonyms is a higher-level cognitive task, workers take longer to do word similarity tasks than image verification and sentiment analysis tasks. We manage a speedup of $6.23\times$.

Finally, for topic detection, workers spend significant time analyzing articles in the conventional setting (14.33s on average). With 3 workers, the conventional approach takes 43s. In comparison, our technique delegates 2s for each article. With a redundancy of only 2 workers, we achieve a precision of 0.95, similar to the 0.96 achieved by the conventional approach. The total worker time to label one article using our technique is 4s, a speedup of $10.75\times$.

The mean TLX workload for the control condition was 58.5 ($\sigma=9.3$), and 62.4 ($\sigma=18.5$) for our technique. Unexpectedly, the difference between conditions was not significant ($t(99)=-0.53, p=0.59$). The “temporal demand” scale item appeared to be elevated for our technique (61.1 vs. 70.0), but this difference was not significant ($t(99)=-0.76, p=0.45$). We conclude that our technique can be used to scale crowdsourcing on a variety of tasks without statistically increasing worker workload.

\section{Study 3: Multi-class Classification}
In this study, we extend our technique from binary to multi-class classification to capture an even larger set of crowdsourcing tasks. We use our technique to create a dataset where each image is classified into one category (``people,'' ``dog,'' ``horse,'' ``cat,'' etc.). We compare our technique with a conventional technique~\cite{deng2009imagenet} that collects binary labels for each image for every single possible class.

\textit{Method.} Our aim is to classify a dataset of 2,000 images with 10 categories where each category contains between 100 to 250 examples. We compared three methods of multi-class classification: (1) a \textit{naive approach} that collected 10 binary labels (one for each class) for each image, (2) a \textit{baseline approach} that used our interface and classified images one class (chosen randomly) at a time, and (3) a \textit{class-optimized approach} that used our interface to classify images starting from the class with the most examples. When using our interface, we broke tasks into streams of 100 images displayed for 100ms each. We used a redundancy of 3 workers for the conventional interface and 5 workers for our interface. We calculated the precision and recall scores across each of these three methods as well as the cost (in seconds) of each method.

\textit{Results.} (1) In the \textit{naive approach}, we need to collect 20,000 binary labels that take 1.7s each. With 5 workers, this takes 102,000s (\$170 at a wage rate of \$6/hr) with an average precision of $0.99$ and recall of $0.95$. (2) Using the \textit{baseline approach}, it takes 12,342s (\$20.57) with an average precision of $0.98$ and recall of $0.83$. This shows that the baseline approach achieves a speedup of $8.26\times$ when compared with the naive approach. (3) Finally, the \textit{class-optimized approach} is able to detect the most common class first and hence reduces the number of times an image is sent through our interface. It takes 11,700s (\$19.50) with an average precision of $0.98$ and recall of $0.83$. The class-optimized approach achieves a speedup of $8.7\times$ when compared to the naive approach. While the speedup between the baseline and the class-optimized methods is small, it would be increased on a larger dataset with more classes.

\section{Application: Building ImageNet}
Our method can be combined with existing techniques~\cite{deng2014scalable, song2011contextualizing, parkash2012attributes, biswas2013simultaneous} that optimize binary verification and multi-class classification by preprocessing data or using active learning. One such method~\cite{deng2014scalable} annotated ImageNet (a popular large dataset for image classification) effectively with a useful insight: they realized that its classes could be grouped together into higher semantic concepts. For example, ``dog,'' ``rabbit'' and ``cat'' could be grouped into the concept ``animal.'' By utilizing the hierarchy of labels that is specific to this task, they were able to preprocess and reduce the number of labels needed to classify all images. As a case study, we combine our technique with their insight and evaluate the speedup in collecting a subset of ImageNet.

\textit{Method.} We focused on a subset of the dataset with 20,000 images and classified them into 200 classes. We conducted this case study by comparing three ways of collecting labels: (1) The naive approach asked 200 binary questions for each image in the subset, where each question asked if the image belonged to one of the 200 classes. We used a redundancy of 3 workers for this task. (2) The optimal-labeling method used the insight to reduce the number of labels by utilizing the hierarchy of image classes.  (3) The combined approach used our technique for multi-class classification combined with the hierarchy insight to reduce the number of labels collected. We used a redundancy of 5 workers for this technique with tasks of 100 images displayed at 250ms.

\textit{Results.} (1) Using the naive approach, this would result in asking 4 million binary verification questions. Given that each binary label takes 1.7s (Table~\ref{tab:precision_recall_speedup}), we estimate that the total time to label the entire dataset would take 6.8 million seconds (\$11,333 at a wage rate of \$6/hr). (2) The optimal-labeling method is estimated to take 1.13 million seconds (\$1,888)~\cite{deng2014scalable}. (3) Combining the hierarchical questions with our interface, we annotate the subset in 136,800s (\$228).  We achieve a precision of $0.97$ with a recall of $0.82$. By combining our $8\times$ speedup with the $6\times$ speedup from intelligent question selection, we achieve a $50\times$ speedup in total.

\newpage
\section{Discussion}
We focused our technique on positively identifying concepts. We then also test its effectiveness at classifying the absence of a concept. Instead of asking workers to react when they see a ``dog,'' if we ask them to react when they do \textit{not} see a ``dog,'' our technique performs poorly. At $100$ms, we find that workers achieve a recall of only $0.31$, which is much lower than a recall of $0.94$ when detecting the presence of ``dog''s. To improve recall to $0.90$, we must slow down the feed to $500$ms. Our technique achieves a speedup of $2\times$ with this speed. We conclude that our technique performs poorly for anomaly detection tasks, where the presence of a concept is common but its absence, an anomaly, is rare. More generally, this exercise suggests that some cognitive tasks are less robust to rapid judgments. Preattentive processing can help us find ``dog''s, but ensuring that there is no ``dog'' requires a linear scan of the entire image.


To better understand the active mechanism behind our technique, we turn to concept typicality. A recent study~\cite{iordan2015basic} used fMRIs to measure humans' recognition speed for different object categories, finding that images of most typical examplars from a class were recognized faster than the least typical categories. They calculated typicality scores for a set of image classes based on how quickly humans recognized them.
In our image verification task, $72\%$ of false negatives were also atypical. 
Not detecting atypical images might lead to the curation of image datasets that are biased towards more common categories. For example, when curating a dataset of dogs, our technique would be more likely to find usual breeds like ``dalmatians'' and ``labradors'' and miss rare breeds like ``romagnolos'' and ``otterhounds.'' More generally, this approach may amplify biases and minimize clarity on edge cases. Slowing down the feed reduces atypical false negatives, resulting in a smaller speedup but with a higher recall for atypical images. 


\section{Conclusion}
We have suggested that crowdsourcing can speed up labeling by encouraging a small amount of error rather than forcing workers to avoid it. We introduce a rapid slideshow interface where items are shown too quickly for workers to get all items correct. We algorithmically model worker errors and recover their intended labels. This interface can be used for binary verification tasks like image verification, sentiment analysis, word similarity and topic detection, achieving speedups of $10.2\times$, $10.2\times$, $6.23\times$ and $10.75\times$ respectively. It can also extend to multi-class classification and achieve a speedup of $8.26\times$. Our approach is only one possible interface instantiation of the concept of encouraging some error; we suggest that future work may investigate many others. Speeding up crowdsourcing enables us to build larger datasets to empower scientific insights and industry practice. For many labeling goals, this technique can be used to construct datasets that are an order of magnitude larger without increasing cost.

\section{Acknowledgements}
This work was supported by a National Science Foundation award 1351131. Thanks to Joy Kim, Juho Kim, Geza Kovacs, Niloufar Salehi and reviewers for their feedback.

\balance{}
\bibliographystyle{SIGCHI-Reference-Format}
\bibliography{references}

\end{document}